\documentclass[reprint,aps,prb,twocolumn,superscriptaddress]{revtex4-2}  

\usepackage{natbib,bm,array,bbm,physics,hyperref}
\usepackage{mathtools, blkarray, mathrsfs}
\usepackage{amsmath,amsfonts,amssymb}
\usepackage{outlines, cancel}

\usepackage{diagbox} 
\usepackage[caption=false]{subfig}
\hypersetup{colorlinks = true,allcolors = blue }
\usepackage{graphicx}

\usepackage{multirow, outlines}
\usepackage{environ} 
\usepackage{tikz}

\usetikzlibrary{matrix,fit}
\usepackage[boxed,ruled,vlined]{algorithm2e}


\usepackage{graphicx,epsf}
\usepackage{color}
\usepackage{xcolor}

\newcommand{\bseq}{\begin{subequations}}
\newcommand{\eseq}{\end{subequations}}
\newcommand{\bsplit}{\begin{split}}
\newcommand{\esplit}{\end{split}}

\NewEnviron{eqsplit}{%
\begin{equation}\begin{split}
  \BODY
\end{split}\end{equation}
}

\usepackage{adjustbox}
\usepackage[boxed,ruled,vlined]{algorithm2e}

\begin{document}

\title{Classical Criticality via Quantum Annealing}
\author{Pratik Sathe}
\affiliation{Theoretical Division, Quantum \& Condensed Matter Physics, Los Alamos National Laboratory}
\affiliation{Information Science \& Technology Institute, Los Alamos National Laboratory, Los Alamos, NM 87545, USA}
\author{Andrew D. King}
\affiliation{D-Wave Quantum Inc., Burnaby BCV5G 4M9, Canada}
\author{Susan M. Mniszewski}
\affiliation{Computer, Computational and Statistical Sciences (CCS) Division (CCS-3 Information Sciences), Los Alamos National Laboratory, Los Alamos, NM 87545, USA}
\author{Carleton Coffrin}
\affiliation{Advanced Network Science Initiative, Los Alamos National Laboratory, Los Alamos, NM 87545, USA}
\author{Cristiano Nisoli}
\affiliation{Theoretical Division, Quantum \& Condensed Matter Physics, Los Alamos National Laboratory}
\author{Francesco Caravelli}
\affiliation{Theoretical Division, Quantum \& Condensed Matter Physics, Los Alamos National Laboratory}

\date{\today}
\begin{abstract}
Quantum annealing provides a powerful platform for simulating magnetic materials and realizing statistical physics models, presenting a compelling alternative to classical Monte Carlo methods.
We demonstrate that quantum annealers can accurately reproduce phase diagrams and simulate critical phenomena without suffering from the critical slowing down that often affects classical algorithms. 
To illustrate this, we study the piled-up dominoes model, 
which interpolates between the ferromagnetic 2D Ising model and Villain's fully frustrated ``odd model''.
We map out its phase diagram and for the first time, employ finite-size scaling and Binder cumulants on a quantum annealer to study critical exponents for thermal phase transitions.
Our method achieves systematic temperature control by tuning the energy scale of the Hamiltonian, eliminating the need to adjust the physical temperature of the quantum hardware.
This work demonstrates how, through fine-tuning and calibration, a quantum annealer can be employed to apply sophisticated finite-size scaling techniques from statistical mechanics.
Our results establish quantum annealers as robust statistical physics simulators, offering a novel pathway for studying phase transitions and critical behavior.
\end{abstract}
\maketitle

\newpage

Quantum simulation--- the realization of models of collective, interacting qubits within a quantum machine--- has been a central motivation in the development of quantum computing and quantum information science~\cite{LloydUniversal1996, feynman2018simulating, GeorgescuQuantum2014}.
Numerous digital as well as analog quantum simulation approaches have been implemented successfully over the last few decades~\cite{DaleyPractical2022}.
Among them, quantum annealing—an analog method grounded in the adiabatic theorem—was originally introduced to solve classical combinatorial optimization problems~\cite{FinnilaQuantum1994,KadowakiQuantum1998,SantoroTheory2002,AlbashAdiabatic2018,YarkoniQuantum2022,Tasseffemerging2024}.
However, when repurposed for analog simulation, superconducting qubit-based quantum annealing (QA) has shown promise in simulating a range of quantum condensed matter systems~\cite{HarrisPhase2018, KingObservation2018,NishimuraGriffithsMcCoy2020,WeinbergScaling2020,KingQuantum2021,KingCoherent2022,RajakQuantum2023,KingQuantum2023,NarasimhanSimulating2024,AliQuantum2024}--- in some cases, potentially outperforming the best-known classical algorithms~\cite{KingScaling2021,KingBeyondclassical2025} (see also Refs.~\cite{MauronChallenging2025,TindallDynamics2025} for critical perspectives).

While quantum simulation applications of QA have received significant attention (particularly in the so-called \textit{coherent regime} in which coupling to the environment is negligible~\cite{KingCoherent2022,KingQuantum2023,AliQuantum2024}), we propose and demonstrate that QA can be used to simulate thermodynamic phase transitions in \textit{classical} statistical physics models when operated in the \textit{incoherent} regime~\footnote{The ``incoherent regime'' usually corresponds to long annealing times, of the order of tens to hundreds of microseconds}, thus leveraging a perceived drawback as an advantage.
In this regime, quantum annealers have been shown to sample from the equilibrium (i.e., Gibbs or Boltzmann) distributions corresponding to the programmed classical Ising models, due to a combination of non-adiabaticity and environmental interactions~\cite{DumoulinChallenges2014, AminSearching2015,BenedettiEstimation2016,BenedettiQuantumAssisted2017}.

A central goal in the study of statistical physics models is the identification of critical points and their associated critical exponents. In the absence of exact solutions--- which is often the case--- Markov-Chain Monte Carlo (MCMC) methods are a standard numerical approach~\cite{BinderMonte2010,NewmanMonte1999,landau2021guide}.
These algorithms aim to generate and analyze samples from equilibrium distributions across different temperatures and parameter regimes.
Given its ability to perform a similar sampling task, QA offers a natural alternative for simulating classical systems. 

QA has previously been employed to study various aspects of classical statistical physics models, particularly those involving geometric frustration.
Examples include reproducing the ground-state ($T=0$) phase diagram of the (classical) Shastry-Sutherland model~\cite{KairysSimulating2020}; measuring order parameters as functions of tuning parameters at an unknown but fixed temperature~\cite{ParkFrustrated2022,MarinModeling2024}; investigating artificial spin-ice dynamics~\cite{KingQubit2021,Lopez-BezanillaKagome2023, KingMagnetic2023,Lopez-BezanillaQuantum2024}; and extracting critical exponents in classical 3D spin-glass models under parameter variation at fixed temperature~\cite{HarrisPhase2018}.
In some of these studies, Gibbs sampling is not explicitly invoked, and the focus is either on zero-temperature behavior or on sampling with model parameters varied at a fixed, device-dictated temperature.
\begin{figure*}[ht]
  \centering
  \subfloat[]{
  \includegraphics[width=0.20\linewidth]{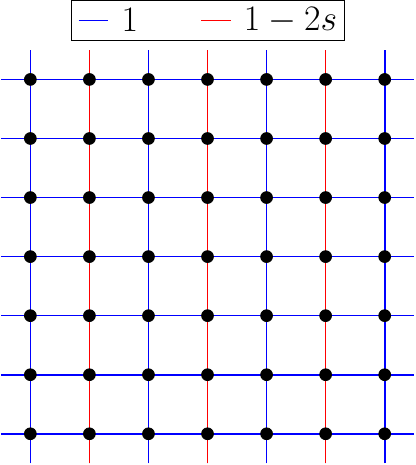}
  \label{fig:lattice}
  }\hfill 
  \subfloat[]{%
    \includegraphics[width=0.2275\textwidth]{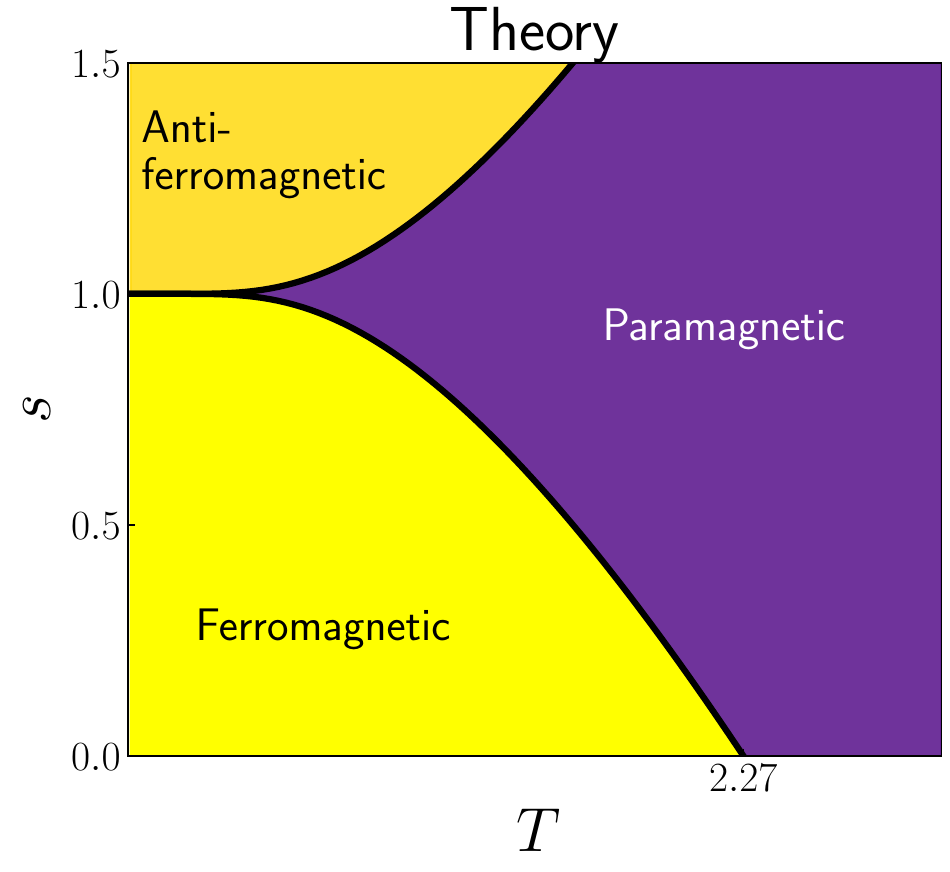}%
    \label{fig:exact_solution_phase_diagram}
  } \hfill 
  \subfloat[]{%
    \includegraphics[width=0.26\textwidth]{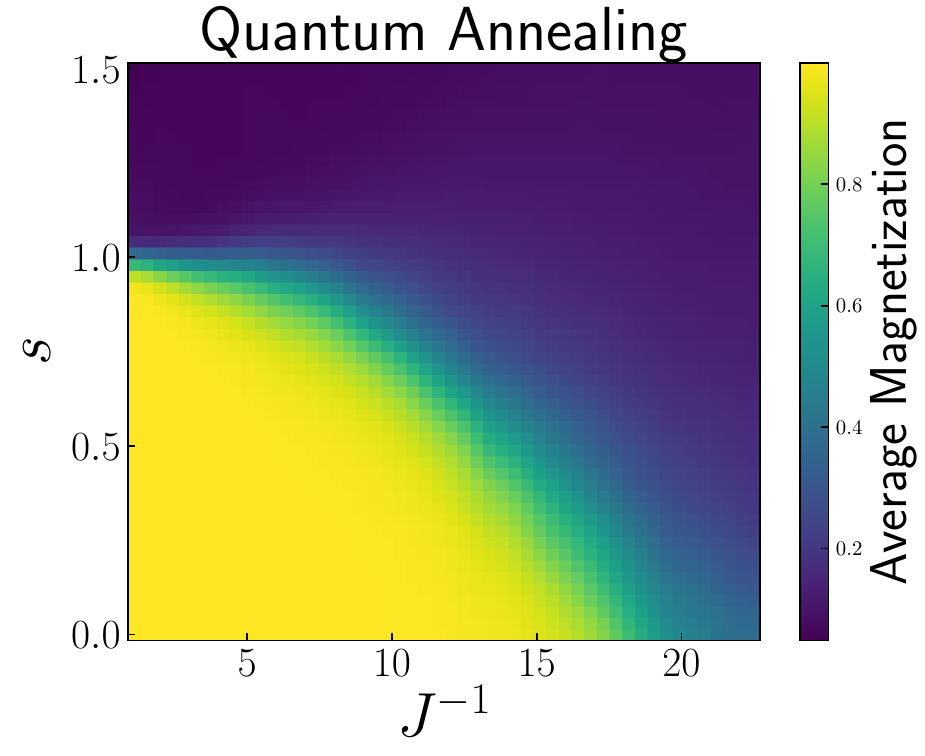}%
    \label{fig:dwave_144_ferro}
  }
  \subfloat[]{%
    \includegraphics[width=0.26\textwidth]{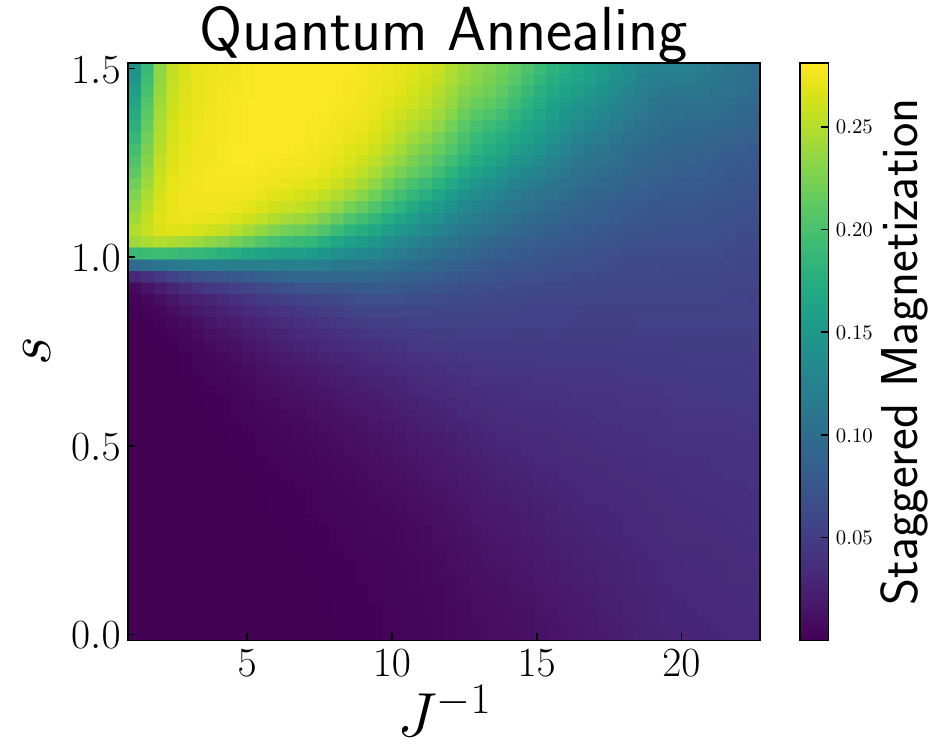}%
    \label{fig:dwave_144_antiferro}
  } 
  \caption{(a) The PUD model is an Ising model with a tunable degree of frustration. Defined on a square lattice, it has two types of nearest-neighbor couplings $J_{ij}$, shown here in blue and red, with the Hamiltonian defined as $H = -\sum_{i<j}J_{ij} s_i s_j$.
  (b) The phase diagram in the $s$-$T$ plane obtained from the exact solution. 
  Data corresponding to quantum annealing samples are shown in (c) and (d).
  Specifically, the ferromagnetic order parameter (sub-figure c) and the antiferromagnetic (or staggered) order parameter (sub-figure d) corresponding to samples generated using D-Wave's \texttt{Advantage2\_prototype2.6} device for a toroidal system of size $12\times 12$ are in agreement with the exact solution.}
  \label{fig:main}
\end{figure*}

Quantum annealing implementations often deviate from ideal Boltzmann sampling~\cite{RaymondGlobal2016}, exhibiting issues such as biased sampling within degenerate ground states~\cite{MatsudaGroundstate2009,BoixoExperimental2013,MandraExponentially2017}, the influence of noise~\cite{VuffrayProgrammable2022} and quantum effects~\cite{MorrellSignatures2023}. 
Although high-quality Gibbs sampling has been demonstrated under specific conditions~\cite{NelsonHighQuality2022} and can be improved through specialized techniques~\cite{MarshallPower2019, SandtEfficient2023}, the identification of the precise conditions when that occurs--- and achieving fine, systematic control over the sampling temperature--- remains computationally expensive and generally infeasible for the large system sizes typically studied in statistical physics.
Hence, despite its promise, QA has yet to be used to reconstruct full thermodynamic phase diagrams with thermal phase transitions without privileged access to the device's physical temperature.

In contrast, we show that the sampling temperature can be systematically controlled by tuning the Hamiltonian's energy scale.
Despite known challenges in achieving high-quality Boltzmann sampling with QA, we demonstrate that QA can still capture thermodynamic phase transitions in classical statistical models and serve as a viable alternative to traditional Monte Carlo methods. 
To this end, we present a methodology for reconstructing phase diagrams and critical behavior in classical systems, which we apply to the Piled-Up Dominoes (PUD) model~\cite{AndreFrustration1979}—-- a classically solvable spin system with tunable frustration. 
In general, the competing and mutually-incompatible interactions that define geometric frustration result in the emergence of exotic phases such as spin ice and spin liquid phases~\cite{MoessnerGeometrical2006}.
Furthermore, numerical simulations of frustrated systems are also usually harder~\cite{BinderMonte2010}.
The PUD model thus provides a meaningful and challenging benchmark for assessing the effectiveness of our approach.

\section{The Piled Up Domino Model}
The PUD model is a family of Hamiltonians interpolating between Villain's fully-frustrated ``odd model'' (which does not have a phase transition)~\cite{VillainSpin1977}, and the ferromagnetic 2D Ising model (which does have a phase transition)~\cite{OnsagerCrystal1944}, the interpolation being controlled by a parameter $s$:
\bseq \label{eq:OV_model_defn_eqns}
\begin{align}
H(s) &= (1-s) H_{\text{2D-Ising}} + s H_{\text{Villain}}  \label{eq:OV_as_interpolation}\\
&= - \sum_{x,y} \sigma_{x,y} \sigma_{x,y+1} - \sum_{x,y} \sigma_{2x,y} \sigma_{2x,y+1}  \nonumber \\
 & - (1-2s) \sum_{x,y} \sigma_{2x+1,y} \sigma_{2x+1,y+1}, \label{eq:OV_model_defn_si_sj}
\end{align}
\eseq
where $\sigma_{x,y}=\pm 1$ denotes a (classical) Ising spin located at position $(x,y)$ on a square grid.

When visualized over a 2D grid (see Fig.~\ref{fig:lattice}), all the horizontal links and the links on alternate vertical columns have a coupling strength of $1$ (FM), with all the remaining vertical columns having coupling values equal to $1-2s$~\footnote{
We note that the sign convention for the Hamiltonian in terms of the exchange energy $J$ in Eq.~\eqref{eq:OV_model_defn_eqns}, while standard in condensed matter physics, is the opposite of the one used in D-Wave's QA convention.}.
The model therefore has a tunable frustration. 
It has been solved exactly using transfer matrix methods~\cite{AndreFrustration1979} and more recently via a dimer mapping in Ref.~\cite{Caravelliexactly2022} (wherein it was referred to as the Onsager-Villain model), thus providing a good testbed to probe the potential of QA in the study of criticality.
Its rich phase diagram includes ferromagnetic, paramagnetic, and antiferromagnetic phases, separated by critical lines in the $s$-$T$ (where $T$ is temperature) plane (Fig.~\ref{fig:exact_solution_phase_diagram}). 

To control the sampling temperature, we consider a simple model of D-Wave's QA devices in which the user inputs a Hamiltonian $H_\text{input}$, and the device samples from the corresponding Boltzmann distribution at a fixed, but unknown, device-dependent temperature $T_\text{sampler}$.
That is, the probability of generating a spin sample $\{\sigma_{x,y}\}$ is given by
\begin{align}
    p(\{\sigma_{x,y}\}) = \frac{1}{\mathcal Z} e^{-\beta_\text{sampler} H_\text{input}(\{\sigma_{x,y}\})}, \label{eq:sampling_probability_H_input}
\end{align}
where $\mathcal Z$ is the normalizing factor (also known as the partition function), and $\beta_\text{sampler}=1/T_\text{sampler}$.
To generate samples for an $H$ [such as the PUD model from Eq.~\eqref{eq:OV_model_defn_si_sj} at some value of $s$ of interest], we set $H_\text{input} = J H$ for some value of ``energy scale'' $J$.
It follows from Eq.~\eqref{eq:sampling_probability_H_input} that the generated samples correspond to the Hamiltonian $H$, but at inverse temperature $\beta_\text{effective}=J \beta_\text{sampler}$, or equivalently,
\begin{align}
    T_\text{effective} \propto J^{-1}. \label{eq:temp_propto_inverse_energy_scale}
\end{align}
Thus, increasing the inverse energy scale of the input Hamiltonian proportionately increases the sampling temperature of the device.
For convenience, we will henceforth drop the subscript `effective' while referring to the effective temperature and its inverse.

\begin{figure}
      \includegraphics[width=0.96\linewidth]{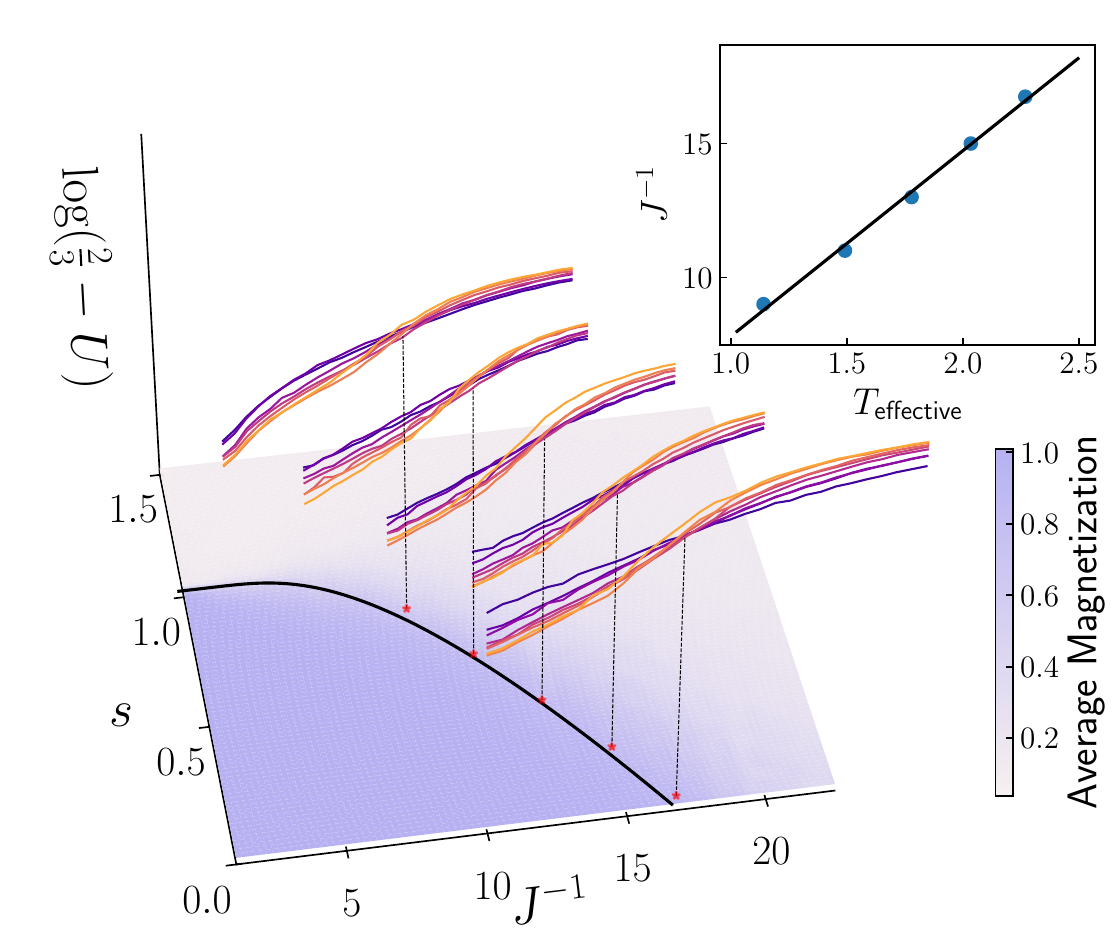}
       \caption{The critical inverse energy scale $J_c^{-1}$ is located for 5 values of $s$ using the intersection of Binder cumulant curves for various system sizes.
       Using the exact solution $T_c(s)$ and $J_c^{-1}(s)$ extracted from QA samples, we plot $J_c^{-1}$ as a function of the effective sampling temperature $T$. 
       We find that $J_c^{-1}$ is linearly proportional to $T$, consistent with the initial assumption Eq.~\eqref{eq:temp_propto_inverse_energy_scale} (see inset). 
       In the main figure, we show the ferromagnetic order parameter (i.e., $\langle \abs{m} \rangle$) for a toroidal system of size $12\times 12$, as a function of $s$ and $J^{-1}$, along with the $J_c^{-1}$ extracted from five different values of $s$. 
       $J_c^{-1}(s)$ is inferred from the linear fit and the analytical expression for the critical line, and clearly separates the ferromagnetic phase from the paramagnetic phase.}
       \label{fig:3d_fig}
\end{figure}

We note that the dependence of the effective temperature on the energy scale and annealing time have been studied previously in Ref.~\cite{NelsonHighQuality2022}. The inverse energy scale was used as a proxy for the effective temperature in Ref.~\cite{Lopez-BezanillaQuantum2024}. 
However, prior studies either focused on small system sizes or assumed the validity of Eq.~\eqref{eq:temp_propto_inverse_energy_scale} without thorough verification.
In this work, using the exact solution of PUD and the finite-size scaling method, we demonstrate the validity of Eq.~\eqref{eq:temp_propto_inverse_energy_scale} across a range of system sizes (see Fig.~\ref{fig:3d_fig}; details provided below).

The exact solution of the PUD model~\cite{AndreFrustration1979,Caravelliexactly2022} yields 
an $s$-$T$ phase diagram (see Fig.~\ref{fig:exact_solution_phase_diagram}) that exhibits three distinct phases--- ferromagnetic, antiferromagnetic and paramagnetic--- separated by critical lines that emanate from the point $(s=1, T=0)$.

To reproduce this phase diagram using quantum annealers, we implement sampling experiments over a grid of values of $s$ (by programmatically controlling the relative coupler values) and over a grid of $T$ values [by tuning the energy scale of the Hamiltonian, as described in Eq.~\eqref{eq:temp_propto_inverse_energy_scale}].

Using D-Wave's \texttt{Advantage2\_prototype2.6} quantum annealer, we successfully observe all three phases.
(See the Methods section for details of the procedure.) 
Specifically, we generate samples of the model defined on a $12\times 12$ torus over a grid of $(s,J^{-1}$) values and plot the corresponding ferromagnetic order parameter (Fig.~\ref{fig:dwave_144_ferro}), defined as 
\begin{align}
    m = \frac{1}{N} \abs{\sum_{i=1}^N \sigma_i}, \label{eq:mag_op}
\end{align}
and the antiferromagnetic order parameter or staggered magnetization (Fig.~\ref{fig:dwave_144_antiferro}), defined as 
\begin{align}
    m_\text{AFM} = \frac{1}{N} \abs{\sum_{i=1}^N (-1)^{x_i + y_i} \sigma_i}. \label{eq:afm_op}
\end{align}
Here, $N$ denotes the number of spins in the system ($144$ in this case), and $(x_i,y_i)$ denotes the location of the $i^{th}$ spin.
Although thermodynamic phase transitions are typically sharp only in the infinite-size limit, finite-size systems exhibit a smooth transition in the order parameter when crossing a critical point.
We find that the boundary separating the ferromagnetic phase (order parameter close to 1) from the paramagnetic phase (order parameter close to 0) qualitatively matches the critical line obtained from the exact solution; similarly, good agreement is found for the antiferromagnetic-to-paramagnetic critical line.

While computing the order parameters for a finite-sized system does provide a qualitative understanding of the location of critical lines, a more robust method involves the computation of Binder cumulants, which we describe below.

\begin{figure*}[t]
  \centering 
  {
  \includegraphics[width=0.45\linewidth]{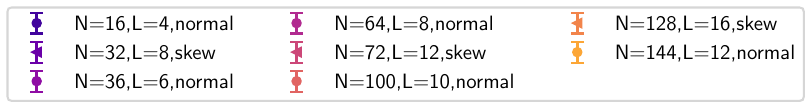}
  } \\
  \subfloat[]{
  \includegraphics[width=0.32\linewidth]{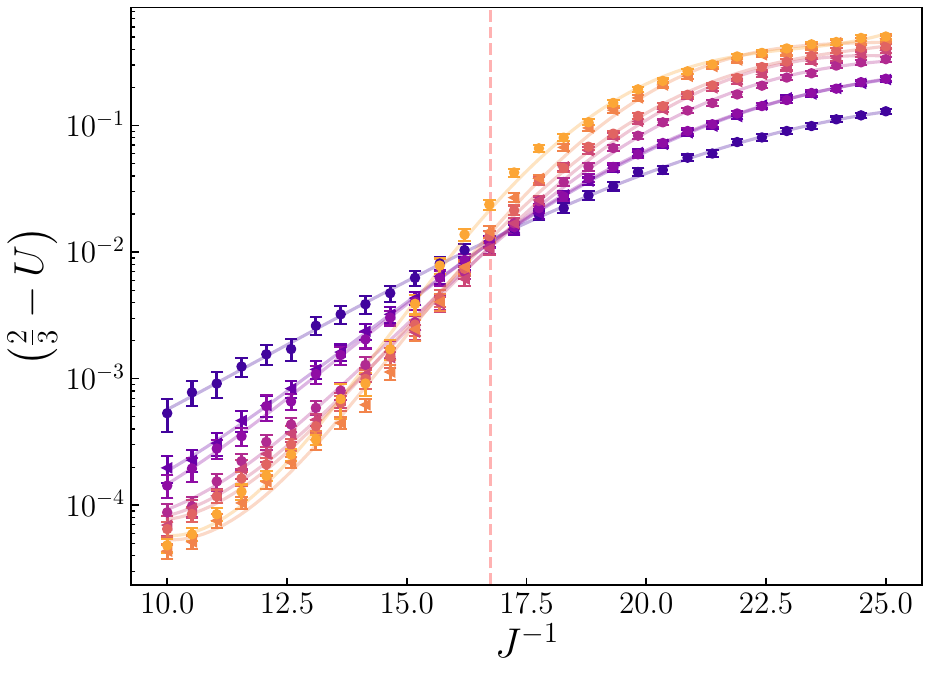}
  \label{fig:binder_cumulants_0.000}
  } 
  \subfloat[]{%
    \includegraphics[width=0.32\textwidth]{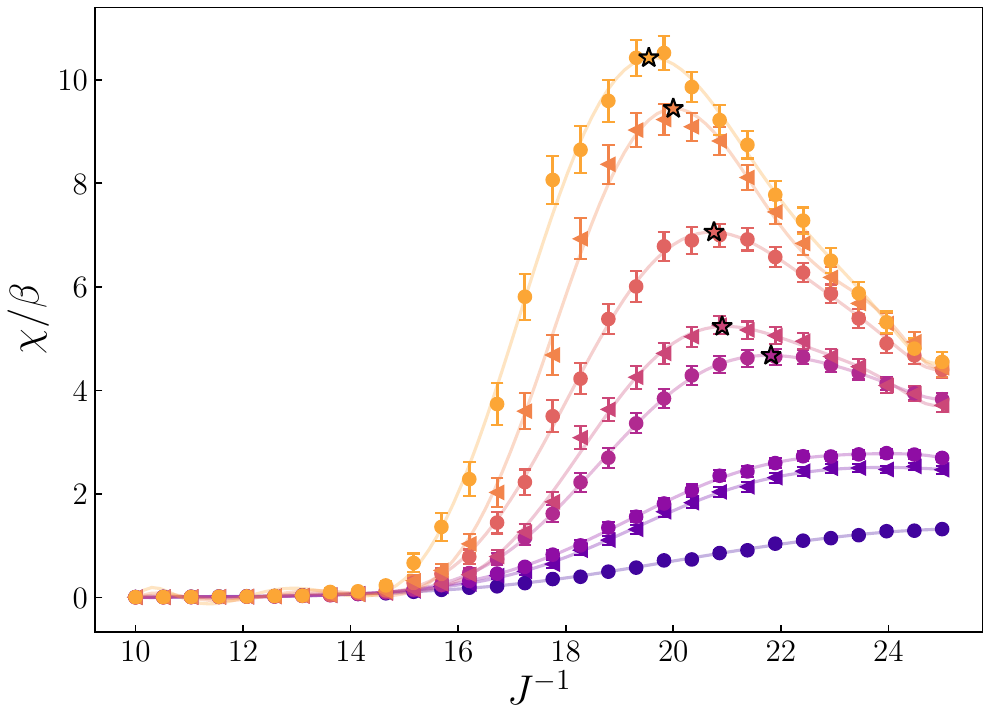}%
    \label{fig:s=0.000_chi_by_beta}
  }  
  \subfloat[]{%
    \includegraphics[width=0.31\textwidth]{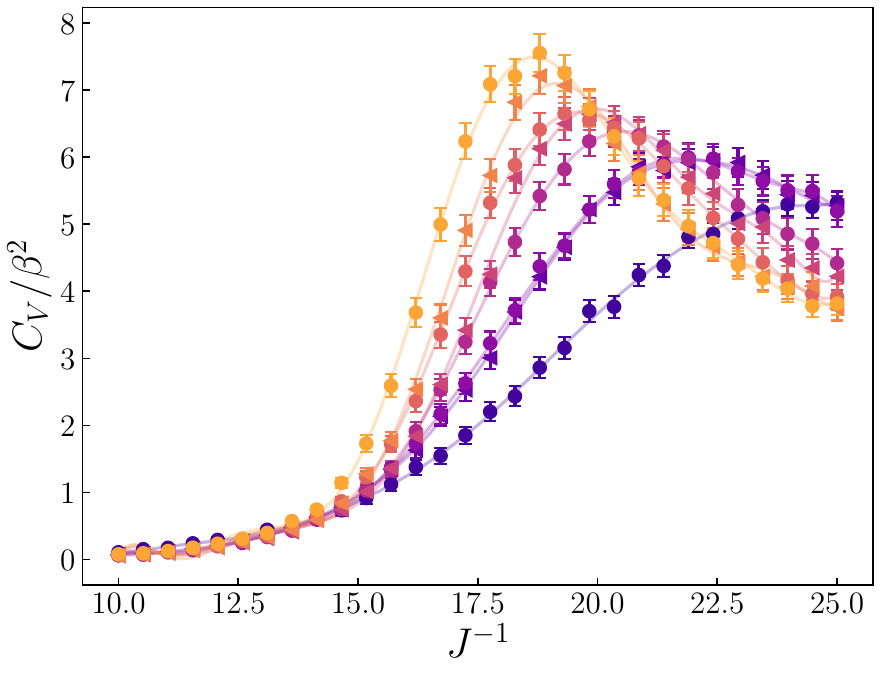}%
    \label{fig:s=0.000_cv_by_beta_squared}
  } 
  \caption{Values of various thermodynamic quantities obtained using quantum annealing samples for various system sizes, for $s=0$. 
  Shown here are (a) the fourth-order Binder cumulants (subtracted from $2/3$), (b) magnetization susceptibility times temperature, $\chi/\beta$ and (c) the heat capacity times temperature squared $C_V/\beta^2$ for various system sizes as a function of the inverse energy scale $J^{-1}$. The star symbols in Fig.~(b) denote the peak values of a polynomial fit for each of the different system sizes.}
  \label{fig:binder_chi_and_cv}
\end{figure*}

\section{Finite-size scaling and critical exponents}
The exact solution indicates that for each value of $s$, the system undergoes a second-order phase transition at a corresponding critical temperature $T_c(s)$. 
(Exact solution expressions for the critical temperature are provided in the Methods section.)
To experimentally identify the critical points as a function of the tuning parameter $s$, we employ two complementary methods: the crossing of Binder cumulants and finite-size scaling (FSS) analysis of susceptibility~\cite{BinderFinite1981}.
We restrict our analysis to the regime $s<1$ (and thus to the ferromagnetic-paramagnetic transition) and extract critical energy scales (i.e., those corresponding to critical temperatures) specifically at $s = 0,0.2,0.4,0.6$ and $0.8$.
(A similar analysis can be implemented for $s>1$, by replacing the ferromagnetic order parameter $m$ by the antiferromagnetic order parameter $m_\text{AFM}$, in the analysis below.)
For each value of $s$, to probe different effective temperatures, we consider a uniform grid of 30  $J^{-1}$ values, centered around the anticipated critical value of $J^{-1}$.
To minimize boundary effects, we implement sampling only for systems with periodic boundary conditions along both spatial directions.
Specifically, we considered toroidal lattices with $N = L \times L$ spins and $L \in \{6,8,10,12\}$, as well as skew-toroidal lattices with $N = (L \times L) /2$ spins and $L \in \{8,12,16\}$ (more details are provided in the Methods section). 
For each system size and each $J^{-1}$ value, we sampled $10,000$ spin configurations and computed the fourth-order Binder cumulant~\cite{BinderFinite1981}
\begin{align}
    U = 1 - \frac{\langle m^4\rangle}{ 3 \langle m^2 \rangle ^2},
\end{align}
where $m$ denotes the magnetization per spin of a configuration. 
While implementing Monte Carlo simulations, the critical temperature can be determined by identifying the intersection of Binder cumulant curves for various system sizes~\cite{BinderFinite1981,BinderMonte2010}.
Analogously, for each value of $s$ considered in our study, we observe a clear crossing of the Binder cumulant curves, enabling us to identify the energy scale $J_c$ corresponding to the critical temperature $T_c$.
For example, Fig.~\ref{fig:binder_cumulants_0.000} shows the Binder cumulant curves for $s=0$.
A slight scatter in intersection points is observed, consistent with finite-size deviations expected to arise from the limited system sizes imposed by hardware constraints.

For each value of $s$, we compare the inverse critical energy scale $J_c^{-1}(s)$ extracted from the crossing of Binder cumulant curves in the quantum annealing data, with the critical temperature $T_c(s)$ obtained from the exact solution.
Consistent with our initial hypothesis, Eq.~\eqref{eq:temp_propto_inverse_energy_scale}, the two quantities exhibit a linear relationship (see inset of Fig.~\ref{fig:3d_fig}), implying self-consistency.
Using the corresponding linear fit and the analytical expression for the critical line $s(T)$ (see Appendix~\ref{sec:exact_sol}), we obtain a prediction for the critical inverse energy scale $J_c^{-1}(s)$ as a function of $s$ for $s\in [0,1]$.
When this curve is overlaid on the average magnetization plot for the $12\times 12$ system size, it delineates the boundary between the ferromagnetic and paramagnetic phases (Fig.~\ref{fig:3d_fig}).

\begin{figure*}[ht]
  \centering
  \subfloat[]{
  \includegraphics[width=0.33\linewidth]{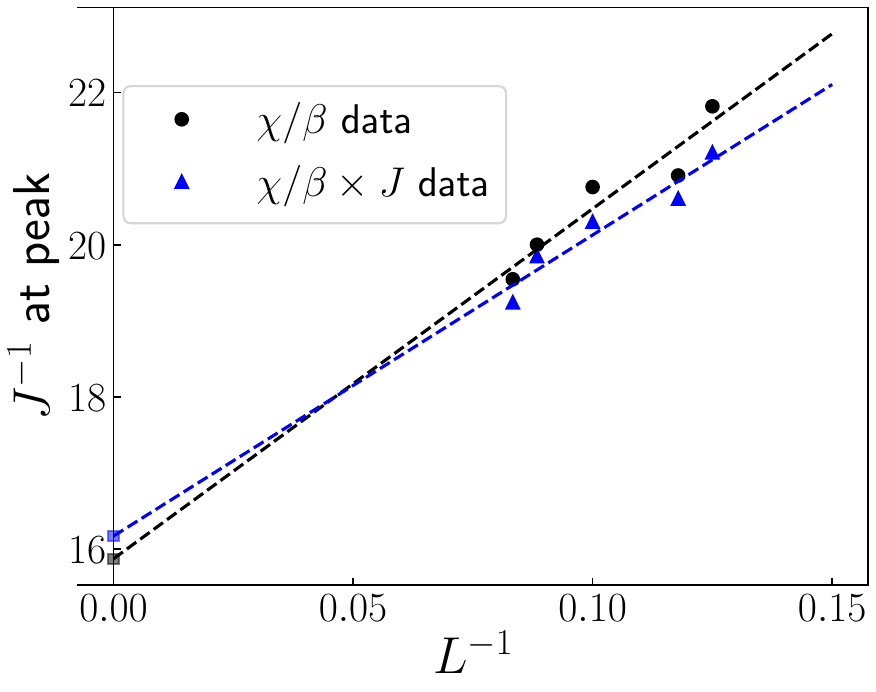}
  \label{fig:intercepts_s=0}
  } \hspace{30pt} 
  \subfloat[]{%
    \includegraphics[width=0.40\textwidth]{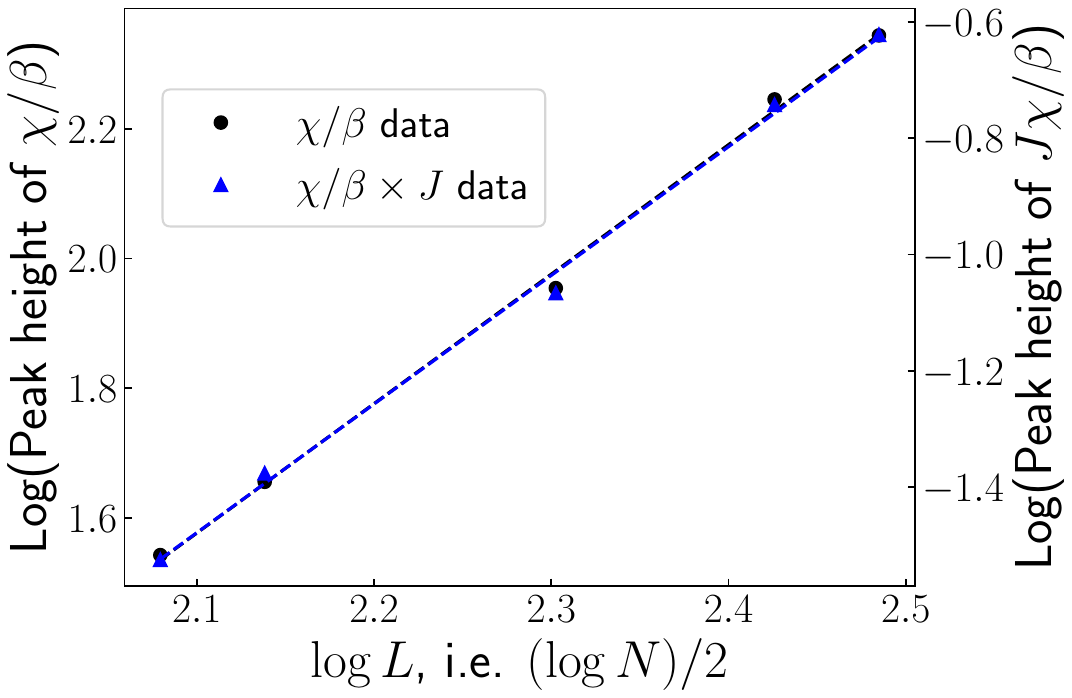}%
    \label{fig:gamma_by_nu_at_s=0}
  } \\
  \subfloat[]{%
    \includegraphics[width=0.35\textwidth]{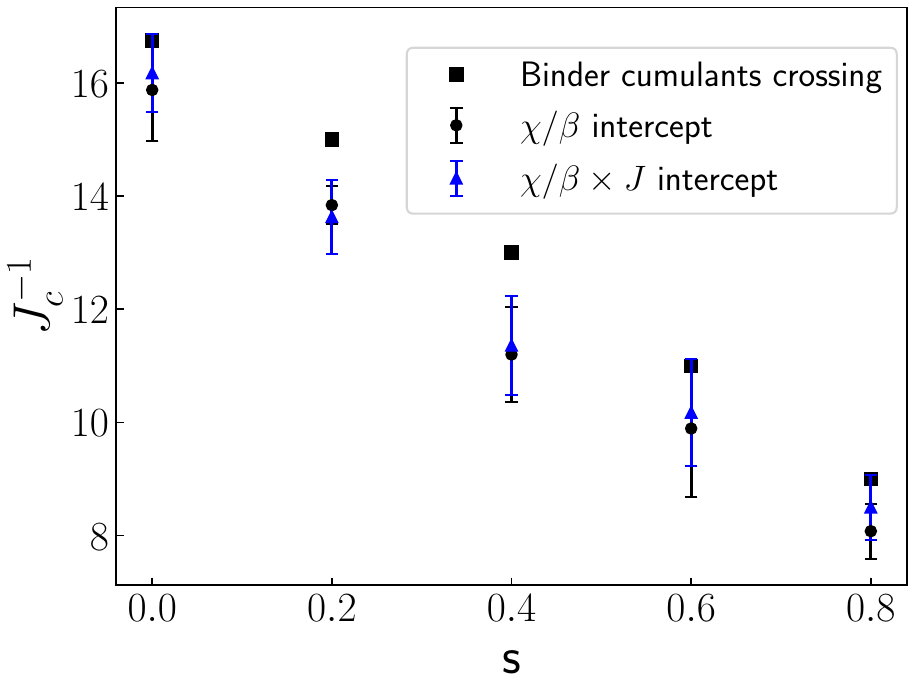}%
    \label{fig:intercept_as_a_function_of_s}
  } \hspace{30pt}
  \subfloat[]{%
    \includegraphics[width=0.35\textwidth]{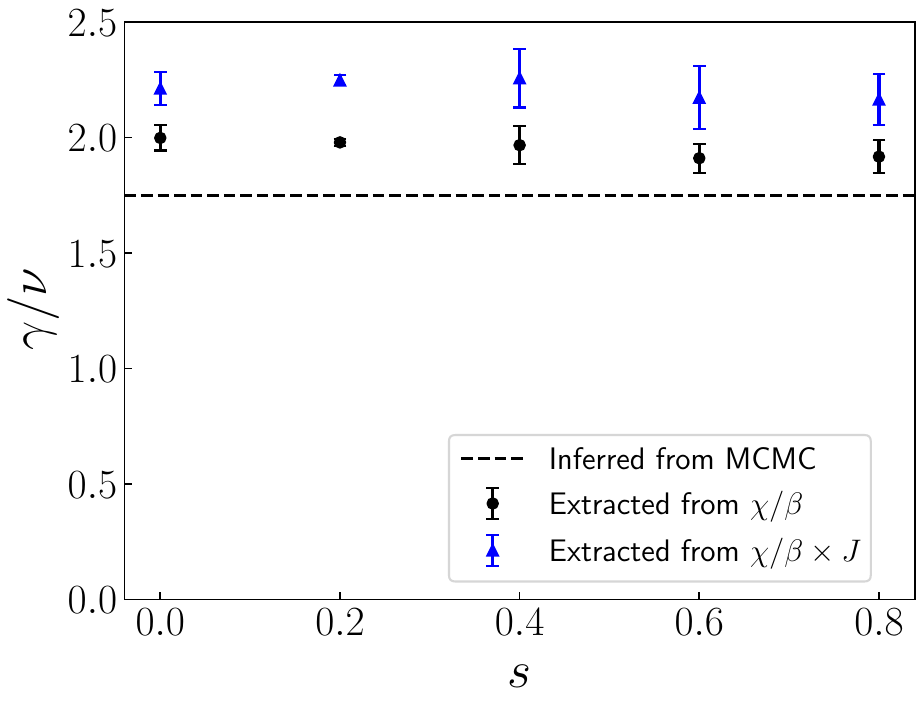}%
    \label{fig:gamma_by_nu_as_a_function_of_s}
  } 
  \caption{Finite size scaling (FSS) analysis on quantum annealer data, for obtaining the critical inverse energy scale $J^{-1}_c$ and the ratio of exponents $\gamma/\nu$.
  Plots (a) and (b) correspond to $s=0$.
  (a) The locations of peaks of $\chi/\beta$ and $\chi/\beta \times J$ as a function of inverse system size $L^{-1}$.
  The intercept corresponding to the best linear fit yields $J_c^{-1}$.
  (b) The slope of the best linear fit of the logarithm of the peaks of $\chi/\beta$ (and of $\chi/\beta\times J$) vs the logarithm of system size $L$ yields $\gamma/\nu$.
  (c) The critical inverse energy scale and (d) the extracted values of $\gamma/\nu$ as a function of the interpolation parameter $s$. 
  The `correct value' (inferred from MCMC), $\gamma/\nu=1.75$ is plotted as a dashed horizontal line.}
  \label{fig:one_by_nu_and_gamma_by_nu}
\end{figure*}

The critical energy scale can also be identified using an FSS analysis of the magnetization susceptibility (per spin) $\chi$ and the heat capacity (per spin) $C_V$ as a function of the system size and energy scale.
To that end, we extract $\chi/\beta$ and $C_V/\beta^2$ from the quantum annealing samples, using 
\bseq
\begin{align}
    \frac{\chi}{\beta} &= N (\langle m^2 \rangle - \langle m \rangle ^2), \label{eq:chi_by_beta_formula} \\
    \text{and }\frac{C_V}{\beta^2} &= \frac{(\langle E^2 \rangle - \langle E \rangle ^2)}{N},
\end{align}
\eseq
where $m$ and $E$ denote the magnetization per spin and energy [with respect to Eq.~\eqref{eq:OV_as_interpolation}, without any energy scaling] of spin configurations.
We note that since $\beta$ is unknown, $\chi$ and $C_V$ cannot directly be extracted by multiplying $\frac{\chi}{\beta}$ and $\frac{C_V}{\beta^2}$ respectively by appropriate powers of $\beta$.
(While $\chi$ was heuristically obtained in Ref.~\cite{HarrisPhase2018} using a time-varying ramp for the longitudinal field, our approach is based on Boltzmann sampling. Hence, using \eqref{eq:chi_by_beta_formula} is more appropriate.)
We plot both these quantities as a function of $J^{-1}$ for various system sizes and for various $s$.
The obtained trends (see Figs.~\ref{fig:s=0.000_chi_by_beta} and \ref{fig:s=0.000_cv_by_beta_squared}) are consistent with those usually observed in Monte Carlo simulations: As we increase the system size, the heights of the peaks increase, and their corresponding temperatures move towards lower temperatures (which correspond to lower values of $J^{-1}$ in our case).

Next, we implement an FSS analysis of the susceptibility as a function of $J^{-1}$ and system size.
We recall that close to a continuous phase transition, the correlation length $\xi$ and the magnetization susceptibility $\chi$ diverge as the reduced temperature $t=(T-T_c)/T_c$ approaches $0$. The critical exponents $\nu$ and $\gamma$ are defined through the relationships
\begin{align}
    \xi \propto t^{-\gamma} \text{ and } \chi \propto t^{-\nu}.
\end{align}
At $s=0$, the exact values of the exponents are known to be $\gamma=1.75$ and $\nu=1$. 
Using the Metropolis Markov-chain Monte Carlo (MCMC) method, we find that the exponents do not change as a function of $s$ for $s \in [0,1)$ (see Methods section for more details about the MCMC simulations). 

We now discuss the procedure for extracting the exponents and critical inverse energy scale using quantum annealing.
To that end, we first recall that the values $\gamma(s)$, $\nu(s)$ and $T_c(s)$ can be obtained from the locations and heights of peaks of the $\chi(L,T,s)$ curves as a function of $T$ for various $L$, with $\chi$ extracted from any sampling method (such as QA or MCMC).
For notational convenience, we will drop the dependence on $s$.
Let $\chi_L$ and $T_L$ denote the height and temperature corresponding to the peak of the curve $\chi(L,T)$ as a function of $T$.
The finite-size scaling ansatz implies that~\cite{NewmanMonte1999,BinderMonte2010}
\bseq \label{eq:FSS_usual}
\begin{align}
    T_L &= T_c (1 + a L^{-1/\nu}), \label{eq:T_c_intercept} \\
    \text{and }\log \chi_L &= b + \gamma/\nu \log L. \label{eq:gamma_by_nu_main}
\end{align}
\eseq
Here, $a$ and $b$ are some constants that are not of interest.
While Eqs.~\eqref{eq:FSS_usual} can be used to extract $\nu$, $\gamma$ and $T_C$ when using MCMC, it is not directly suitable for our quantum annealing-based sampling method, since we do not have direct access to $T$, or to $\chi$. 
Since $T\propto J^{-1}$, we expect that the susceptibility $\chi$ will be proportional to the product $\frac{\chi}{\beta} \times J$. 
Hence, we plot $\frac{\chi}{\beta}\times J$ as a function of $J^{-1}$ for various system sizes, to mimic the dependence of $\chi$ as a function of temperature.
Let us denote the peak heights and locations by $\chi_L$ and $J^{-1}_L$ respectively.
Then, from Eq.~\eqref{eq:temp_propto_inverse_energy_scale} and Eq.~\eqref{eq:T_c_intercept}, it follows that
\begin{align}
    J_L^{-1} &= J_c^{-1} (1 + x_0 L^{-1/\nu}) \label{eq:Jinv_L_susc}.
\end{align}
While a curve fitting procedure is typically used to simultaneously extract $\nu$ and the critical temperature (instead, $J_c^{-1}$ in our case), our quantum-annealing data was too noisy to accurately extract $\nu$. 
Hence, we instead assume $\nu=1$ for all $s$ (validated by MCMC; see Methods section for more details), and plot $J_L^{-1}$ as a function of $L^{-1}$. 
From Eq.~\eqref{eq:Jinv_L_susc} (with $\nu=1$), the intercept then corresponds to $J_c^{-1}$.
The linear fit for $s=0$ is shown in Fig.~\ref{fig:intercepts_s=0}.
The vertical axis intercept yields our estimate for $J_c^{-1}$ at $s=0$.
While these values differ from those obtained using crossings of Binder cumulants, both data sets follow similar trends as a function of $s$.

Turning to Eq.~\eqref{eq:gamma_by_nu_main}, we note that the slope in the linear relationship between $\log \chi_L$ and $\log L$ corresponds to $\gamma/\nu$.
The extracted values of $\gamma/\beta$ and their associated errors (equal to the standard deviation of the slope in the linear fits) are plotted as a function of $s$ in Fig.~\ref{fig:gamma_by_nu_as_a_function_of_s}.
The values are consistently seen to lie around $2.2$, which is off by approximately $25\%$ from the correct value of $1.75$ (inferred from MCMC).
This discrepancy may stem from deviations from the hypothesized relation, Eq.~\eqref{eq:temp_propto_inverse_energy_scale}, which was applied at two stages in the analysis as well as from the limitations imposed by small system sizes.

To avoid these issues, we propose and implement an alternate procedure, based on directly using the peak locations and heights of $\chi/\beta$ (without multiplying by $J$).
This avoids the second potential source of error, i.e., the application of the temperature-inverse energy scale relationship.
Specifically, we propose that the values $\chi_L$, $T_L$ and hence $J_L^{-1}$ in Eqs.~\eqref{eq:FSS_usual} and Eq.~\eqref{eq:Jinv_L_susc} may be replaced by the peaks of $\chi/\beta$ instead, to obtain $\gamma/\nu$ and critical values of inverse energy scales.
We base this on the difference between the peak locations (peak heights) of the two quantities being infinitesimally small (related by a constant) in the limit of infinite system sizes. 
(See the Methods section for an argument.)
While the system sizes in our analysis are fairly small, we find the extracted values of $\gamma/\nu$ to be around $2$ for various values of $s$, thus being closer to the correct answer of $1.75$ (see Fig.~\ref{fig:gamma_by_nu_as_a_function_of_s}).

In contrast to the order parameters [Eqs.~\eqref{eq:mag_op} and~\eqref{eq:afm_op}] discussed above, the susceptibility, heat capacity, and Binder cumulants require the computation of up to fourth-order moments of magnetization and energy, and are thus sensitive to experimental errors.
We find that a calibration-refinement technique is crucial to obtain reliable results~\cite{ChernTutorial2023a}. We discuss it in detail in the Methods section.

\begin{figure}[t]
    \centering
  \includegraphics[width=0.75\linewidth]{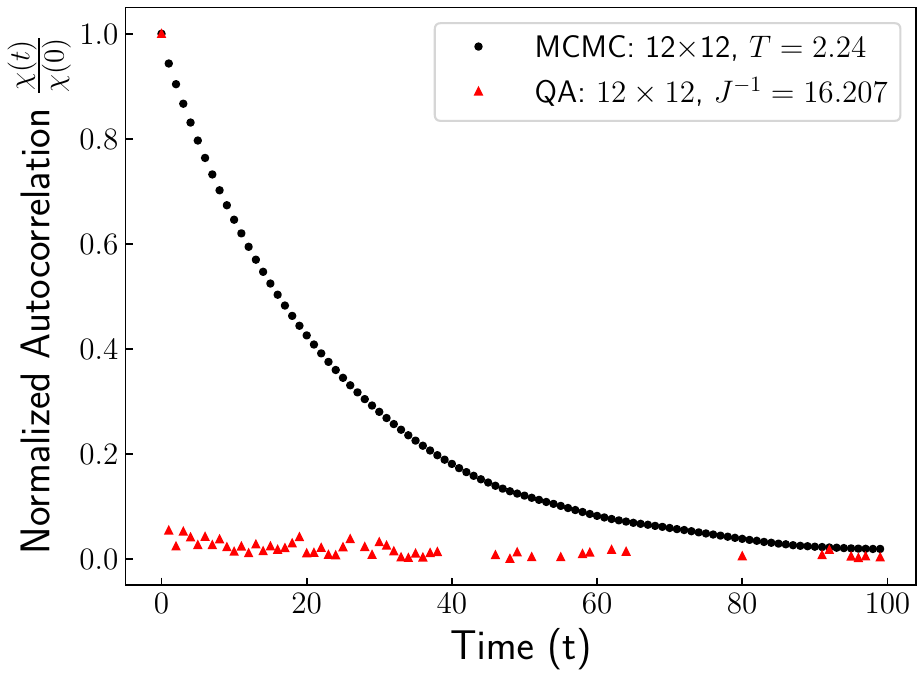}
  \caption{Normalized autocorrelation function $\chi(t)/\chi(0)$ as a function of time $t$ for Markov-chain Monte Carlo (MCMC) simulations and for quantum annealing samples.
  Plotted data corresponds to $s=0.0$. 
  The temperature and inverse energy scale were chosen to be close to their respective critical values.}
  \label{fig:autocorr_comparison}
\end{figure}

\section{QA Circumvents Critical Slowing Down}
Our work addresses a persistent challenge in computational statistical mechanics.
One of the key challenges faced by many Markov-Chain Monte Carlo (MCMC) methods—such as the single-spin flip algorithm—is the phenomenon of critical slowing down near continuous phase transitions~\cite{HohenbergTheory1977,NewmanMonte1999}.
In this regime, the system's convergence to equilibrium becomes significantly slower, and many updates are required to generate statistically independent samples.
This issue is particularly severe for certain frustrated systems such as spin-glasses or spin ices.
While model-specific Monte Carlo algorithms incorporating cooperative updates can partially alleviate the problem~\cite{BinderMonte2010,melko2001long,chern2013degeneracy,macdonald2011classical}, no universal strategy has been found to eliminate it across arbitrary models.

Our QA-based approach naturally evades this issue.
Each sample in QA is generated by initializing the qubits in the all-up state with respect to the Pauli $X$ basis, evolving them with a time-evolving Hamiltonian, and finally measuring in the Pauli $Z$ basis.
Since the initialization operation is identical for each sample (and independent of the measurement output of any prior sample), we expect our samples to be statistically independent by design.
Any residual correlations, if present, would primarily arise from device-level non-idealities rather than the sampling procedure itself.

While critical slowing becomes increasingly significant for large system sizes, our results show that quantum annealing largely circumvents this issue.
To directly compare slowing down for our single spin-flip Metropolis implementation and our quantum annealing approach, we choose the largest system size we used for QA, which is $N=12\times 12$ spins.
For a fair comparison, we implement MCMC simulations for the same system size.
To measure the degree of correlations between samples that are $t$ steps apart, we compute the autocorrelation function $\chi(t)$ using~\cite{NewmanMonte1999}
\begin{align}
    \begin{split}
    \chi(t) &= \frac{1}{t_\text{max}-t} \sum_{t'=0}^{t_\text{max}-t}m(t')m(t'+t) \\
    &- \frac{1}{t_\text{max}-t} \sum_{t'=0}^{t_\text{max}-t} m(t') \times \frac{1}{t_\text{max}-t}\sum_{t'=0}^{t_\text{max}-t} m(t'+t).
    \end{split}
\end{align}
where $t_\text{max}$ is the total number of samples or steps generated by the algorithm. 
(For the quantum annealing simulations, we generated $t_\text{max}=10,000$ samples.)
For the MCMC simulations, following standard convention, one unit of time is taken to correspond to one MCMC update per spin in the system.

In Fig.~\ref{fig:autocorr_comparison}, we plot the normalized autocorrelation function for QA, along with MCMC simulations for $s=0$ close to the critical point. 
The MCMC data exhibits the expected exponential decay in autocorrelation, indicative of correlations persisting over many time steps, especially close to criticality. 
In contrast, the QA result shows a small, approximately constant autocorrelation value, suggesting little to no temporal correlation between samples.
In typical MCMC simulations, the normalized autocorrelation function decays exponentially with a characteristic timescale $T_A$, which diverges with increasing system size near the critical point~\cite{NewmanMonte1999}.
The absence of such behavior in the QA results—particularly the lack of any discernible correlation timescale even near the critical point—supports the conclusion that QA effectively bypasses critical slowing down.

To further investigate correlations near the phase transitions, we examine the average value of the normalized autocorrelation function and again observe no significant increase in its magnitude close to the transition points. (See Methods section for more details.)

We note that the critical slowing down discussed here pertains specifically to the generation of classical samples--- whether via classical MCMC or using QA--- used to study a classical model (such as the PUD model).
In the context of QA, a distinct form of slowing down also arises: the time-dependent Hamiltonian of the QA device typically traverses a quantum phase transition, near which the quantum dynamics of the qubits can slow down or freeze~\cite{HohenbergTheory1977}.
Related phenomena such as the Kibble-Zurek scaling of defects have been probed in quantum annealers~\cite{KingCoherent2022,KingQuantum2023,AliQuantum2024}. 
However, our focus here is not on this quantum slowing down, but rather on the behavior of QA viewed as a sampling algorithm for classical statistical physics models.

\section{Discussion and Outlook}
We have tested quantum annealing (QA) as a compelling alternative to Metropolis Monte Carlo methods for investigating a rich model exhibiting tunable frustration. 
Our analysis successfully demonstrates the extraction of order parameters and the accurate determination of the phase diagram.
Crucially, for the first time, we have implemented the sophisticated and delicate methodology of finite-size scaling within a quantum annealing framework, marking a significant methodological advancement in this field.

Successfully, for each value of $s$  we consistently observe clear intersections of the fourth-order Binder cumulant curves across multiple system sizes, thus corroborating a Gibbs sampling hypothesis. 
This allows us to extract the value of the energy scale that identifies the critical temperature.
We benchmarked our results against exact solution and proved for the first time that the hypothesized relation between the sampling temperature and inverse energy scale is consistent with the critical energy scales obtained using the Binder cumulant analysis.

Using the QA samples, we computed the temperature-scaled heat capacity and magnetization susceptibility across various system sizes.
By employing a finite size scaling analysis of the peaks, we obtained the ratio $\gamma/\nu$ for the ferromagnetic-paramagnetic phase transitions for a range of values of $s$ for $s<1$.
Importantly, we found no evidence of critical slowing down for the QA approach.
We anticipate that with improved hardware quality and denser qubit connectivity--- enabling access to larger system sizes--- our QA based approach can become a viable alternative for classical Monte Carlo algorithms, potentially mitigating the issue of critical slowing down.

\textbf{Acknowledgements}
We thank Cristian Batista for helpful discussions.
The authors acknowledge the support of NNSA for the U.S. DOE at LANL under Contract No. DE-AC52-06NA25396, and Laboratory Directed Research and Development (LDRD) for support through 20240032DR.
Research presented in this article was also supported by the National Security Education Center (NSEC) Informational Science and Technology Institute (ISTI) using the Laboratory Directed Research and Development program of Los Alamos National Laboratory project number 20240479CR-IST.
PS also acknowledges the support via ISTI Fellowship. Assigned: Los Alamos Unclassified Report LA-UR-25-23528.
LANL is managed by Triad National Security, LLC, for the National Nuclear Security Administration of the U.S. DOE under contract 89233218CNA000001. LA-UR-25-23528.

%

\newpage
\clearpage
\appendix 

\twocolumngrid 
\onecolumngrid
\section{Methods}
\subsection{Exact solution: Critical lines in the PUD model} \label{sec:exact_sol}

The study of exactly solvable models in statistical mechanics began with Onsager’s landmark solution of the ferromagnetic Ising model on a two-dimensional square lattice~\cite{OnsagerCrystal1944}. 
Later, Villain introduced a distinct exactly solvable model—the so-called ``odd model” \cite{VillainSpin1977}, now more commonly referred to as the fully frustrated Ising model—where frustration is introduced systematically across the lattice.
Building on these foundational works, Ref.~\cite{AndreFrustration1979} provided the first exact solution of a model that interpolates between Onsager’s and Villain’s limits, which they referred to as the ``Piled-Up Domino Model” (PUD).

Using dimer methods, the model has also been solved more recently in Ref.~\cite{Caravelliexactly2022}. 
Following the Montroll-Potts-Ward method~\cite{montroll} the partition function per site was obtained to be
\begin{align}
    \begin{split}
   \frac{\log Z}{N} &= \log (2\cosh(2J/\kappa T)) \\
   & + \frac{1}{4\pi^2} \int_0^\pi \! d\phi \int_0^\pi \! d\theta \, \log\left[(1+\tilde z^2)^2 - 2\tilde z^2(\cos \phi + \cos \theta)\right],
   \end{split}
\end{align}
where $\tilde z = \tanh\left(2J/\kappa T\right)$, and $\kappa$ is Boltzmann's constant \cite{Caravelliexactly2022}. 
Locations where $Z$ is not analytic can be found, yielding two critical lines: 
(i) for $s\leq1$, a line separating the ferromagnetic (FM) phase from the paramagnetic (PM) phase, and 
(ii) for $s\geq1$, a line separating the antiferromagnetic (AFM) phase from the PM phase.
The FM-PM critical line was found to be given by
\begin{eqnarray}
   s(T)=\frac{1}{2} \left(T \tanh ^{-1}\left(\frac{z^3+z^2+z-1}{z^3+z^2-z+1}\right)+1\right), \label{eq:critline1}
\end{eqnarray}
where $z=\tanh\big(\frac{J}{\kappa T}\big)$. 
At $s=0$, the critical temperature is Onsager's critical temperature $T_c$. 

An alternative, equivalent form of the model’s critical line can be extracted analytically by identifying parameter values at which long-range order vanishes.
This method works for both FM-PM and AFM-PM critical lines.
Defining $J'/J = 1 - 2s$, $K = \beta J$ and $K' = \beta J'$ with $\beta = 1/(k_B T)$, the two critical lines are given by~\cite{AndreFrustration1979}:
\begin{subequations}
\begin{align}
    \sinh(2K) \sinh(K + K') &= 1, \\
    \text{and }\sinh(2K) \sinh(K + K') &= -1.
\end{align}
\end{subequations}
Substituting $K' = (1 - 2s)K$, we find
\begin{equation}
    \sinh(2K) \sinh(2K(1-s)) = \pm 1. 
\end{equation}
Solving for $s$, we obtain the two critical lines:
\begin{equation}
    s_{\pm}(T) = 1 - \frac{T}{2} \sinh^{-1}\left(\frac{\pm 1}{\sinh(2/T)}\right), \label{eq:critline2}
\end{equation}
where the `$+$' sign corresponds to the FM-PM critical line, and the `$-$' sign corresponds to the AFM-PM phase transition. 

\subsection{Quantum Annealing Protocol}
For all our quantum annealing (QA) experiments reported in the main text, we used D-Wave's \texttt{Advantage2\_prototype2.6} device, accessed via the Leap cloud-computing service provided by D-Wave Quantum Inc.~\cite{LeapService}.
We used an annealing time of $T_A=100\ \mu s$ for each experiment.
To compute the ferromagnetic and antiferromagnetic order parameters reported in Fig.~1 of the main text, we obtained $10,000$ samples, with a programmed time gap known as ``readout thermalization'', of $10,000\ \mu s$ to allow the hardware to cool down after each sampling experiment and to improve result quality.
We used $1,000$ bootstrap resamples with $1,000$ samples each in order to compute the mean and standard deviation of our estimates of the two order parameters, and found the errors to be negligible for all considered values of $s$ and $J^{-1}$.

\begin{figure*}[t]
  \centering
  \subfloat[]{
  \adjustbox{valign=m}{\includegraphics[width=0.24\linewidth]{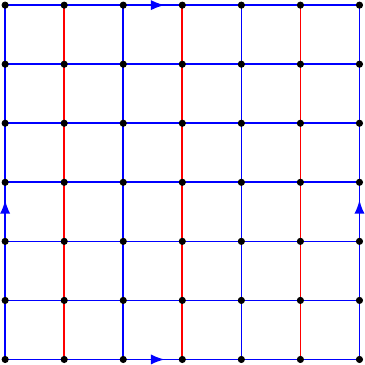}}
  \label{fig:grid_six_by_six}
  }\hfill 
  \subfloat[]{%
    \adjustbox{valign=m}{\includegraphics[width=0.33\textwidth]{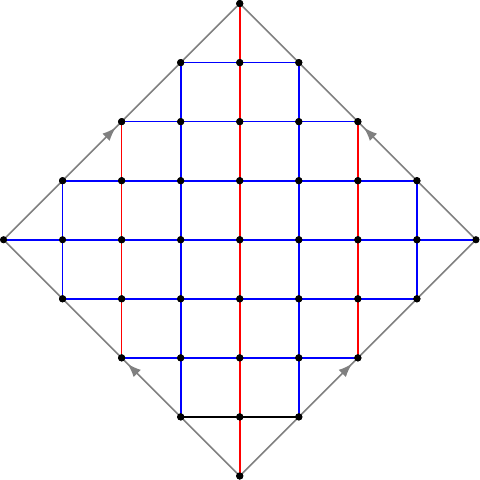}}%
    \label{fig:skew_grid}
  } \hfill
  \subfloat[]{%
    \adjustbox{valign=m}{\includegraphics[width=0.4\textwidth]{skew_embedding_L=12.pdf}}%
    \label{fig:skew_embedding_L=12}
  } 
  \caption{The connectivity graphs for (a) normal and (b) skew toroidal systems for two example system sizes.
  Each spin is shown by a black dot, and red and blue connections correspond to couplings with $J=1-2s$ and $J=s$ values respectively.
  (a) Normal toroidal system corresponding to $L=6$ and number of spins $N=L^2=36$.
  The spins in the top line are identified with the spins in the bottom line, and similarly, the vertical edges on the left and on the right are identified with each other. 
  (b) The connectivity for the skew toroidal system corresponding to $L=8$ with $N=L^2/2=32$ number of spins. 
  The spins on the slanting lines (in gray) are identified (with the orientations specified by the arrows).
  (c) The $6$ disjoint embeddings for the sampling experiments corresponding to system size $N=72$ (skew torus with $L=12$). The gray dots and gray lines denote the nodes and edges of the Zephyr graph connectivity of the \texttt{Advantage2\_prototype2.6} device used for the experiments. 
  The red ($J=1-2s$) and blue ($J=1$) couplers are superimposed on top of the Zephyr graph.}
  \label{fig:appe_1}
\end{figure*}

We extracted the inverse critical energy scales $J_c^{-1}$ and the ratio $\gamma/\nu$ for a grid of interpolation parameter values $s\in \{ 0,0.2,0.4,0.6, 0.8 \}$.
For each $s$ value, $J_c^{-1}$ was obtained by identifying the intersection of Binder cumulants as well as by using finite-size scaling (FSS).
For each $s$, we implemented sampling experiments for many system sizes and for a range of energy scale values.
We implement these experiments for two types of systems: 
(i) normal toroidal systems with number of spins $N=L \times L$, 
and (ii) skew toroidal systems with $N=(L\times L) /2$, with $L$ being an integer.
Graphical representations of the systems corresponding to $N=36$ and $N=32$ are shown in Figs.~\ref{fig:grid_six_by_six} and Fig.~\ref{fig:skew_grid} respectively.

To implement sampling experiments for any system size, the first step is to map—or embed—the toroidal system of interest onto the physical qubits available on the QPU. This mapping must preserve the required connectivity, ensuring that all interacting spins are appropriately connected via couplers. However, the connectivity graph of the problem is often different from that of the QPU hardware. In such cases, it may not be possible to assign each logical spin directly to a physical qubit, and one must instead use chains—groups of connected physical qubits that represent a single logical spin. To minimize errors due to broken chains, we avoid chains altogether in our experiments (by using a chain length of 1 for each qubit), ensuring that each logical spin is mapped to exactly one physical qubit.

Furthermore, to obtain more samples per experiment, for each value of $N$, we used multiple \textit{disjoint} embeddings on the Zephyr graph connectivity of the \texttt{Advantage2\_prototype2.6} device~\cite{BoothbyZephyr}. 
After each anneal, the measurement outputs were processed to extract the corresponding number of spin configurations.
We generated these embeddings for the QPU graph using the \texttt{find\_subgraph} function in the \texttt{minorminer.subgraph} module, applied to different, disjoint regions or subgraphs of the QPU graph.
For example, we generated six disjoint embeddings that were then used for the $N=72$ sampling experiments (see Fig.~\ref{fig:skew_embedding_L=12}).

The various system sizes used and the number of corresponding embeddings are listed in Table~\ref{table:torus_sizes}.
\begin{table}[h] \label{table:torus_sizes}
\centering
\begin{tabular}{|c|c|c|c|}
\hline
Number of spins (N) & Type(Normal/Skew) & L &  Number of embeddings\\
\hline
32 & Skew & 8 & 22 \\
36& Normal & 6 & 16\\
64  & Normal & 8 & 6\\
72 & Skew & 12 & 6\\
100 & Normal & 10 & 2\\
128 & Skew & 16 & 2\\
144 & Normal & 12 & 2\\
\hline
\end{tabular}
\caption{Various system sizes used for the finite-size scaling analysis}
\end{table}

We note that for the PUD model, most couplers are ferromagnetic for any value of $s$. 
(In fact, for $s<0.5$, \textit{all} the couplers are ferromagnetic.)
We found that the quality of the samples was adversely affected by cross-talk and memory effects when most couplers are ferromagnetic, and required a longer pause between consecutive anneals for better result quality.
To reduce these errors, we implemented a gauge transformation $\sigma_{x,y} \rightarrow (-1)^{x+y} \sigma_{x,y}$, which flips the values of all alternate spins.
This results in the sign of each coupler being flipped, $J_{x,y}\rightarrow - J_{x,y}$.
Consequently, the model transforms to one in which all horizontal and alternate vertical columns have \textit{antiferromagnetic} bonds, while all the remaining alternate vertical columns are ferromagnetic, resulting in most bonds being antiferromagnetic.
We obtain the samples generated using QA for the gauge-transformed Hamiltonian, undo the gauge transformation by flipping each measured spin configuration, and then compute the ferromagnetic and antiferromagnetic order parameters.

We note that random gauge transformations have been used in prior works, for example, in Refs.~\cite{MandraExponentially2017,SandtEfficient2023,NelsonHighQuality2022,MarshallPower2019}, to reduce the effects of non-idealities in the hardware.
Unlike these methods, we use a fixed gauge transformation described above, and work with it throughout, since having a fixed gauge also enables a simpler calibration refinement procedure which is described below. 

\subsection{Calibration Refinement}
To reduce the effects of non-idealities in quantum annealing, such as crosstalk, device variation, and environmental noise, we implement a calibration refinement protocol which is also known as ``shimming''~\cite{ChernTutorial2023a}.
This procedure is often crucial to obtain reliable results, and has been used in many previous works that have leveraged quantum annealing to study condensed matter physics models~\cite{KingCoherent2022,KingObservation2018,KingScaling2021}. 
We implemented shimming to extract the crossings of the Binder cumulant and to extract the value of $\gamma/\nu$.

Let us briefly outline the main ideas underlying shimming before providing more details of the procedure.
While sampling from an Ising model that has some symmetries (such as a $\mathbb{Z}_2$ symmetry, or more general graphical symmetries), certain statistical quantities such as mean magnetization of individual spins and the so-called frustration probability of couplers (defined below) are expected to satisfy corresponding conditions in an ideal implementation.
During our calibration refinement procedure, we adjust the values of the FBOs (flux-bias offsets) at each individual qubit and coupler strengths (individual values of $J_{i,j}$) by small amounts, to nudge the system to satisfy these symmetries.

The two main diagnostic quantities that we used for the calibration refinement protocol are average magnetization for each spin, $m_i$, and coupler frustration probability $f_{ij}$ for the coupler that connects qubits $i$ and $j$.
These quantities are defined as~\cite{ChernTutorial2023a}
\bseq \label{eqs:defn_magn_and_frust}
\begin{align}
    m_i &= \langle s_i \rangle, \\
    \text{and }f_{ij} &= \frac{1+ \langle s_i s_j \rangle \operatorname{sign} J_{ij}}{2},
\end{align}
\eseq
with $\langle . \rangle$ denoting the average of a quantity of $100$ measurements.
At any value of $s$, the PUD model has an external longitudinal field value of $0$ at each spin.
Consequently, on average, we expect the magnetization $m_i$ (without taking the absolute value) of each spin to be $0$.
Using gradient descent (with step size denoted by $\alpha_\Phi$), we adjust the values of FBOs for each qubit individually, to lead toward an average magnetization of $0$.

On the other hand, we do not have a quantitative prediction of the value of $f_{ij}$ for any coupler. 
Nonetheless, at any value of $s\neq 0$ for a toroidal system, there are three ``coupler orbits''~\cite{ChernTutorial2023a}.
Referring to Fig.~1a, these orbits form 3 sets:
\begin{outline}[enumerate]
    \1 Orbit 1: All horizontal couplers.
    \1 Orbit 2: All vertical couplers belonging to the blue columns.
    \1 Orbit 3: All vertical couplers belonging to the red columns.
\end{outline}
On a toroidal geometry, all couplers within any orbit should have the same frustration probability. 
We use gradient descent (with step size denoted by $\alpha_J$) and adjust the value of each coupler strength to lead the system towards smaller spreads in frustration probability within each coupler orbit.

We implemented calibration refinement separately for each value of $s\in \{0,0.2,0.4,0.6,0.8\}$ and a grid of values of $J^{-1}$.
We further incorporated smoothing of FBOs and coupler strengths across energy scales at each iteration, and an adaptive adjustment of gradient descent step sizes ($\alpha_\Phi$ and $\alpha_J$)~\cite{ChernTutorial2023a}.
The full calibration refinement procedure for a fixed value of $s$ is provided in Procedure~\ref{algo:Calibration_refinement}.

\begin{algorithm}[htbp] \label{algo:Calibration_refinement}
    \caption{Calibration refinement at a fixed $s$ for extracting $\chi/\beta$ and Binder cumulant experiments} \label{algo:gen_Wannier_func}
    \SetKwInOut{Input}{Input}
    \SetKwInOut{Output}{Output}
    \Input{(i) For a value of interest of $s$, a grid of $N_\text{grid}$ values of $J^{-1}$ centered at the anticipated value of the critical energy scale. 
    (ii) A list of values of system sizes $N$.
    (iii) For each $N$, an embedding with $\mathcal M_\text{embd}(N)$ number of disjoint embeddings that maps the model of size $N$ to the QPU graph.}
    \emph{Procedure:} 
    For each of system size $N$, implement the following steps:
    \begin{outline}[enumerate]
    \1 Implement the gauge transformation $s_{x,y} \rightarrow (-1)^{x+y} s_{x,y}$ for each of the $\mathcal M_\text{embd}(N)$ number of disjoint embeddings. All embeddings combined, let us denote the qubits by $q_i$, the FBOs by $\Phi_i$ with $i \in \{1,\dotsc, N \times \mathcal M_\text{embd} \}$.
    \1 Divide the list of couplers into three sets corresponding to the three coupler orbits specified above.
    \1 Choose initial gradient descent step sizes $\alpha_\Phi$ and $\alpha_J$. Initialize FBO values $\Phi_i = 0$ for all $i$.
    \1 For each value of energy scale $J$ in the chosen grid of values, initialize the coupler strengths to be $J_{i,j}=J \times (1-2s)$ or $J_{i,j} = J \times 1$ chosen appropriately.
    \1 \label{emun:loop_1} For each value of energy scale $J$ in the chosen grid of energy scale values, implement the following steps.
        \2 Using current values $\{ J_{ij}\}$ and $\{ \Phi_i \}$ generate 100 samples from quantum annealing. Obtain the average magnetization $m_i$ for each used qubit $i$, and frustration probability $f_{ij}$ for each used coupler $ij$, averaged over the measurements. (See Eqs.~\ref{eqs:defn_magn_and_frust}.)
        \2 For each qubit $i$, set 
        \begin{align}
            \Phi_i \leftarrow \Phi_i - \alpha_\Phi m_i.
        \end{align}
        \2 For each of the three coupler orbits, compute average frustration probabilities within the orbit.
         For each used coupler $i,j$, update
        \begin{align}
            J_{ij} \leftarrow J_{i,j} (1+\alpha_J (f_{ij} - \overline f) ),
        \end{align}
        where $\overline f$ denotes the average frustration probability of the coupler orbit that coupler $(i,j)$ belongs to.
        \2 Update the values of $\alpha_J$ and $\alpha_\Phi$ using the history of values of frustration probabilities and average qubit magnetizations using the adaptive step size procedure outlined in Ref.~\cite{ChernTutorial2023a}.
    \1 For each qubit $i$, implement a moving window (of size 3) average for $\Phi_i$, across the $N_\text{grid}$ number of inverse energy scale values. Obtain updated values of $\Phi_i$s for each energy scale.
    \1 \label{emun:loop_2} Similarly, for each coupler $(i,j)$, implement a moving window (of size 3) average of $J_{ij}$ across the $N_\text{grid}$ number of inverse energy scale values. Obtain updated values of $J_{ij}$s for each energy scale.
    \1 Repeat steps~\ref{emun:loop_1}-\ref{emun:loop_2} $N_\text{shim}$ number of times.
    \end{outline}
    \Output{Coupler and FBO values for all the qubits used for sampling, for all inverse energy scale grid values.}
\end{algorithm} 
We used $N_\text{shim}=1000$ shimming iterations, and initial gradient descent step sizes $\alpha_J = 0.02 \times J$, and $\alpha_\Phi=10^{-6}$.

The values of coupler strengths and FBOs for all the system sizes and the energy scale grid for each $s$ were then used to obtain $10,000$ spin configurations.
Our estimates and corresponding standard deviation values for the Binder cumulant, and temperature-scaled magnetization susceptibility $\chi/\beta$ and heat capacity $C_V/\beta^2$ were computed using 1,000 bootstrap resamples with 1,000 samples each.

\subsection{Finite Size Scaling using temperature-scaled magnetization susceptibility}
The locations (i.e., temperature values) and heights of the peaks of $\chi(L,T)$ for various $L$ can be used to determine $\nu$, $\gamma$ and $T_c$ using a finite size scaling (FSS) analysis.
We claimed in the main text that for large system sizes, it is justifiable to apply identical FSS procedures to the peaks of $\chi/\beta$ instead.
Here, we provide an argument in support of this claim.
(We have dropped the dependence of $s$, since it plays no role below.)

We start with the finite-size scaling ansatz
\begin{align}
    \chi(L,t) &= L^{\gamma/\nu} f (L^{1/\nu} t), \label{eq:FSS_ansatz}
\end{align}
where $t=(T-T_c)/L$ denotes the reduced temperature.

We note that the location of the peak of $\chi/\beta$ for a system size $L$, denoted by $\widetilde T_L$ here, satisfy
\begin{align}
    0 &= \left. \pdv{(\chi/\beta)}{T} \right\vert_{\widetilde T_L} \\
    &= \chi(L,\widetilde t_L) + (1+\widetilde t_L) \left. \dv{\chi(L,t)}{t}\right\vert_{\widetilde T_L}. \label{eq:fss_steps_1}
\end{align}
The location of the peak of $\chi$ for system size $L$, denoted here by $T_L$, satisfies
\begin{align}
    \left. \dv{\chi(L,t)}{t} \right\vert_{T_L} &= 0, \\
    \implies f'(L^{1/\nu} t_L) &= 0. \label{eq:derivative_of_f_is_zero_at_tL}
\end{align}
From Eq.~\eqref{eq:fss_steps_1} and Eq.~\eqref{eq:FSS_ansatz}, we get
\begin{align}
   (1 + \widetilde t_L) &= -\frac{f(L^{1/\nu} \widetilde t_L)}{f'(L^{1/\nu} \widetilde t_L) L^{1/\nu}}
\end{align}
Taylor expanding the numerator as well as the denominator in $\Delta_L t \coloneqq \widetilde t_L - t_L$, we get 
\begin{align}
    (1 + \widetilde t_L) &= - \frac{f(L^{1/\nu} t_L) + f'' (L^{1/\nu} t)\vert_{t_L}  L^{2/\nu}(\Delta_L t)^2 + O(\Delta_L t ^3)}{L^{1/\nu} \Delta_L t f''(L^{1/\nu}t)\vert_{t_L} + O(\Delta_L t ^2)} \frac{1}{L^{1/\nu}}
\end{align}
Here, w used Eq.~\eqref{eq:derivative_of_f_is_zero_at_tL} to eliminate the first order derivative of $f$.
For large $L$, the dominant terms in the numerator and the denominator are the respective first terms.
Therefore, we get
\begin{align}
    (1 + \widetilde t_L) &\approx -\frac{1}{L^{2/\nu}}\frac{f(L^{1/\nu} t_L) }{\Delta_L t f''(L^{1/\nu} t)\vert_{t_L} }\\
    \text{or } \Delta_L t &\approx -\frac{1}{L^{2/\nu}} \frac{1}{1+\widetilde t_L} g(L^{1/\nu} t_L),
\end{align}
where $g(x)= \frac{f(x)}{f''(x)}$. 
We note that $g(L^{1/\nu} t_L)$ is a constant independent of $L$, since it equals the value of $f(x)/f''(x)$ at a fixed $x$ (specifically, the $x$ corresponding to the maximum of $f(x)$).
Since we also expect $\widetilde t_L\approx 0$ for large $L$, we can also ignore the $(1+\widetilde t_L)$ term in the denominator.
Hence, we arrive at the expression
\begin{align}
    \Delta_L t \sim -\frac{1}{L^{2/\nu}},
\end{align}
which approaches $0$ as $L\rightarrow \infty$.

In other words, the location of the peak of $\chi/\beta$ as a function of $T$ is a good approximation for the location of the peak of $\chi$ as a function of $T$ for large system sizes.

Let us now compare the heights of the peaks for the two cases.
Their ratio equals
\begin{align}
    \frac{(\chi/\beta)\vert_{\text{peak}}}{\chi\vert_\text{peak}} &= \frac{T_c(1+\widetilde t_L) \chi(\widetilde t_L, L)}{ \chi(t_L,L)}
\end{align}
To first order in $\Delta_L t$, $\chi(\widetilde t_L, L) = \chi(t_L,L)$ due to Eq.~\eqref{eq:derivative_of_f_is_zero_at_tL}.
Hence, conclude that 
\begin{align}
    \frac{(\chi/\beta)\vert_{\text{peak}}}{\chi\vert_\text{peak}} &\approx T_c (1+\widetilde t_L) \\
    &\approx T_c,
\end{align}
where we used $(1+\widetilde t_L)\approx 1$ for large $L$.
Hence, for large system sizes, we expect that the peak heights of $\chi/\beta$ to be the same as the peak heights of $\chi$, but scaled by a constant factor of $T_c$.

\subsection{Monte Carlo simulations}
We recall that the values of the exponents for the $s=0$ limit of the PUD model (i.e., the ferromagnetic Ising model on the 2d square lattice) are known from the exact solution.
Although the expression for the partition function of the PUD model (in the absence of a longitudinal field) has been obtained via exact solutions for all values of $s$, the critical exponents are unknown for $s\neq 0$, to our knowledge. 

In the main text, we probed the critical exponents $\gamma$ and $\nu$ for the ferromagnetic-paramagnetic phase transitions for $s \in [0,1)$.
To verify the correctness of the QA results, we compute the exponents $\gamma$, $\nu$ and $\beta$ using the Metropolis-Hastings Markov-chain Monte Carlo (MCMC) algorithm, and infer the exponent $\alpha$ using the hyperscaling relation $d\nu=2-\beta$. 

We recall that the exponents are defined via the relations $\xi\propto t^{-\gamma}$, $\chi \propto t^{-\nu}$, $C_V \propto t^{-\alpha}$ and $m\propto t^{\beta}$ as $t\rightarrow 0$, where $t$ denotes the reduced temperature. 

We implemented single-flip MCMC simulations for $s=0,0.1,\dotsc,0.9$ at intervals of $0.1$.
For each value of $s$, we implemented simulations for toroidal systems of size $L\times L$ with $L$ values $10,20,50,100,150,200,250$.
For each simulation, we initialized the system in a random spin configuration partially polarized (at $70\%$) along the $+z$ direction. 
(This was done to avoid getting stuck in the so-called `striped' metastable states, which are known to occur if a random zero magnetization configuration is used as the initial configuration~\cite{KurchanPhase1996, SpirinFreezing2001, SpirinFate2001}.)
Next, we implemented 40,000 steps per spin to let the system equilibrate. 

To obtain estimates of errors in our estimates of various thermodynamic quantities, we first computed the auto-correlation function $\chi(t)$ for each simulation~\cite{NewmanMonte1999}.
The autocorrelation time was obtained using the formula $\tau = \sum_{t=0}^{T} \frac{\chi(t)}{\chi(0)}$.
Standard deviation errors for average magnetization and average energy were then computed using the equation
\begin{align}
    \sigma = \frac{2\tau}{t_\text{max}} \left( \overline{m^2} - \overline{m}^2 \right).
\end{align}
Using the autocorrelation time $\tau$, the number of statistically independent samples is estimated to be $\mathcal{N} / 2\tau$, where $\mathcal N$ denotes the number of Monte Carlo steps per spin for each simulation. 
For all our MCMC simulations, we implemented $\mathcal N=2,000,000$ steps per spin.
To obtain standard deviation errors for the fourth-order Binder cumulant, magnetization susceptibility $\chi$, and heat capacity $C_V$, we used 1000 bootstrap datasets, with each dataset containing $\mathcal N/2\tau$ number of samples. 

\begin{figure*}[htbp]
  \centering
  \subfloat[]{%
    \includegraphics[width=0.32\textwidth]{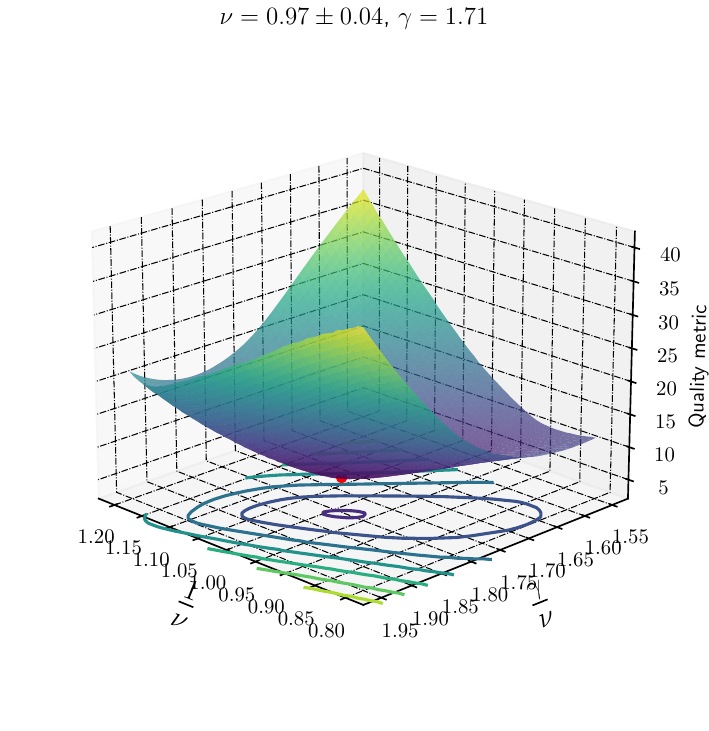}%
    \label{fig:chi_quality_3d_plot_s=0.0}
  }  
  \subfloat[]{
  \includegraphics[width=0.32\linewidth]{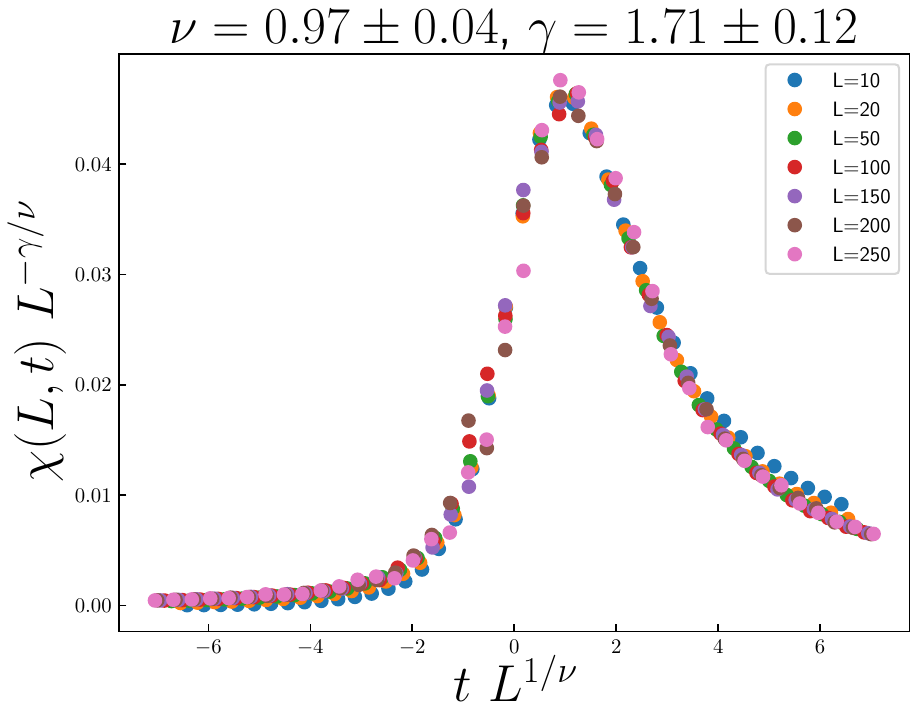}
  \label{fig:chi_data_collapse_extracted_exponents_s=0.0}
  } 
  \subfloat[]{%
    \includegraphics[width=0.32\textwidth]{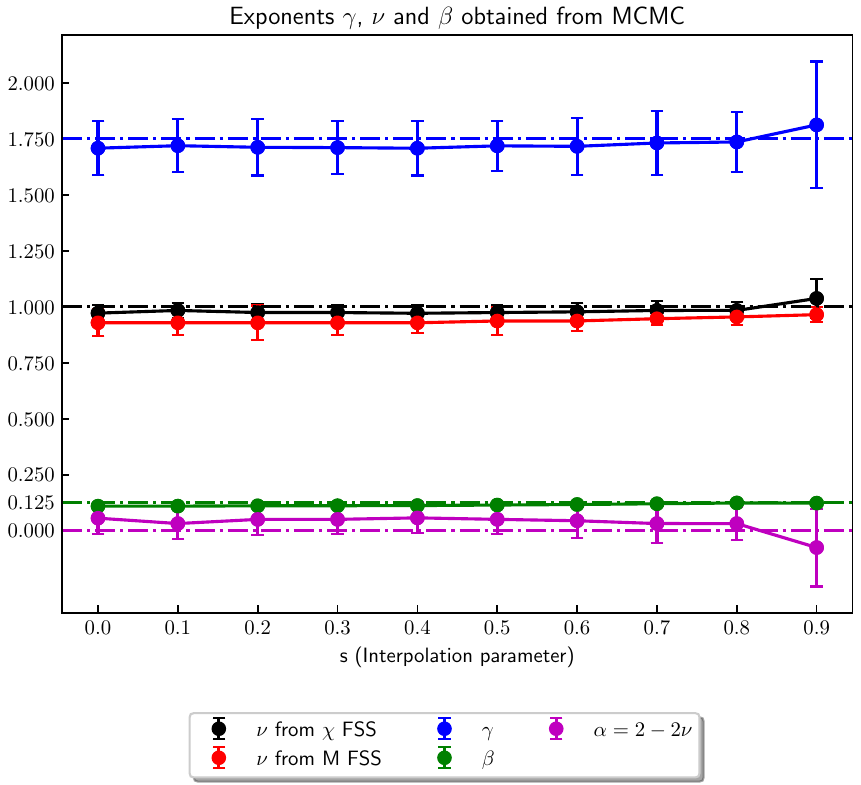}%
    \label{fig:exponents_vs_s_mcmc}
  } 
  \caption{Markov-chain Monte Carlo results. (a) The quality metric, Eq.~\eqref{eq:modified_quality_metric} as a function of $\frac{1}{\nu}$ and $\frac{\gamma}{\nu}$ is minimized at $\frac{1}{\nu}=1.2$ and $\frac{\gamma}{\nu}=1.95$ for $s=0$. (b) The corresponding data collapse for $\chi$ at $s=0.0$. (c) Extracted values of the critical exponents as a function of $s$ obtained using finite size scaling (FSS). The dashed lines indicate the known exact values at $s=0.0$.}
  \label{fig:MCMC_results}
\end{figure*}

To obtain critical exponents, we used the data collapse method.
Specifically, we obtained estimates of $\gamma$ and $\nu$, we plotted $\chi(L,T) L^{-\gamma/\nu}$ vs $tL^{1/\nu}$, where $t=(T-T_c)/T_c$ for various values of $\gamma$ and $\nu$.
The values resulting in the best data collapse were obtained using a modified version of the quality metric described in Ref.~\cite{Bhattacharjeemeasure2001}.
Specifically, we used the quality metric
\begin{align}
    P =  \frac{1}{\mathcal N} \sum_p L_p^{\gamma/\nu}\sum_{j\neq p} \sum_{i,\text{over}} \abs{L_j^{-\gamma/\nu} \chi(L_j,t) - g_p(L^{1/\nu}t)}. \label{eq:modified_quality_metric}
\end{align}
Here, we introduced the normalization factor $L_p^{\gamma/\nu}$ to ensure that indefinitely increasing $\gamma$ does not spuriously result in a smaller quality factor.
The critical exponents and along with their errors, were then obtained by minimizing the quality metric using the Nelder-Mead method.
As a representative example, we show a 3d visualization of the quality metric for the data collapse for $\chi$ at $s=0$ in Fig.~\ref{fig:chi_quality_3d_plot_s=0.0}, and the corresponding data collapse in Fig.~\ref{fig:chi_data_collapse_extracted_exponents_s=0.0}.  

A similar data collapse analysis was implemented for average magnetization, in order to obtain the exponents $\nu$ and $\beta$. To extract the exponent $\alpha$, we used the hyper-scaling relation $d\nu = 2 - \beta$.

A plot of the extracted exponents as a function of $s$ is shown in Fig.~\ref{fig:exponents_vs_s_mcmc}.

\subsection{Critical Slowing Down}
\begin{figure}
    \centering
    \includegraphics[width=0.45\linewidth]{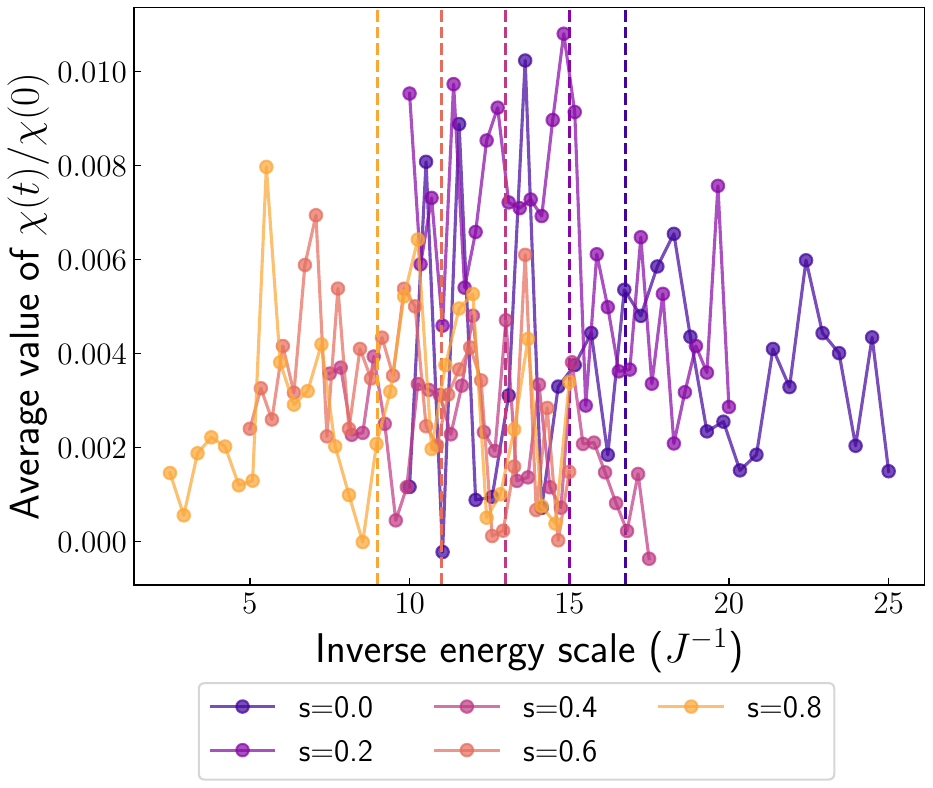}
    \caption{The average value (over 100 steps) of the normalized autocorrelation function, $\chi(t)/\chi(0)$, for quantum annealing samples for various values of $s$, as a function of the inverse energy scale. Vertical lines mark the extracted values of critical inverse energy scale.}
    \label{fig:avg_autocorr_all_s}
\end{figure}

In the main text, we showed that the normalized autocorrelation function, $\chi(t)/\chi(0)$ for the QA samples does not exhibit an exponential drop with increasing $t$, which is typically observed in MCMC simulations.
Instead, at any value of energy scale, the QA data exhibits a roughly constant value of the correlation time for $t>1$.
Critical slowing down can be characterized by the corresponding time scale (called the autocorrelation time) diverging with increasing system size and decreasing distance from the critical temperature.
As an alternative metric, we instead compute the average value of the normalized autocorrelation function over $t\in \{1,\dotsc, 100\}$ for the largest system size. 
Across all values of $s$ examined, this average remains low and shows no noticeable peak near the extracted inverse critical temperature (see Fig.~\ref{fig:avg_autocorr_all_s})—once again indicating the absence of critical slowing down in QA sampling.

\end{document}